\documentclass[aps,twocolumn,groupedaddress,showpacs]{revtex4}
\usepackage{graphicx}
\usepackage{dcolumn}
\usepackage{bm}
\usepackage{epstopdf}
\usepackage[dvipsnames,usenames]{color}
\usepackage{color}
\usepackage{float}
\usepackage{amsmath}
\usepackage{amssymb}
\usepackage{amsthm}
\usepackage{mathrsfs}

\usepackage[T1]{fontenc}
\usepackage[skip=1ex]{caption}
\usepackage{booktabs,tabularx}
\usepackage{siunitx}

\emergencystretch=\maxdimen
\begin{document}
\title{Methodological investigation into the noise influence on nanolasers' large signal modulation}

\author{T.~Wang$^{1,2,\footnote{wangtao@hdu.edu.cn}}$, J.L.~Zou$^{3}$, G.P.~Puccioni$^{4}$, W.S.~Zhao$^{1,2}$, X.~Lin$^{5}$, H.S.~Chen$^{5}$, G.F.~Wang$^{1,2}$, G.L. Lippi$^{6,\footnote{Gian-Luca.Lippi@inphyni.cnrs.fr}}$}

\affiliation{$^1$Engineering Research Center of Smart Microsensors and Microsystems of MOE, Hangzhou Dianzi University, Hangzhou, 310018, China}
\affiliation{$^2$School of Electronics and Information, Hangzhou Dianzi University, Hangzhou, 310018, China}
\affiliation{$^3$School of Communication Engineering, Hangzhou Dianzi University, Hangzhou, 310018, China}
\affiliation{$^4$Istituto dei Sistemi Complessi, CNR, Via Madonna del Piano 10, I-50019 Sesto Fiorentino, Italy}
\affiliation{$^5$Interdisciplinary Center for Quantum Information, College of Information Science and Electronic Engineering, Zhejiang University, Hangzhou 310027, China}
\affiliation{$^6$Universit\'e C\^ote d'Azur, Institut de Physique de Nice (INPHYNI), UMR 7010 CNRS, 1361 Route des Lucioles, F-06560 Valbonne, France}

\date{\today}

\begin{abstract}
Nanolasers are considered ideal candidates for communications and data processing at chip-level thanks to their extremely reduced footprint, low thermal load and potentially outstanding modulation bandwidth, which in some case has been numerically estimated to exceed hundreds of GHz.  The few experimental implementations reported to date, however, have so-far fallen very short of such predictions, whether because of technical difficulties or of overoptimistic numerical results.  We propose a methodology to study the physical characteristics which determine the system's robustness and apply it to a general model, using numerical simulations of large-signal modulation.  
Changing the DC pump values and modulation frequencies, we further investigate the influence of intrinsic noise, considering, in addition, the role of cavity losses.  Our results confirm that significant modulation bandwidths can be achieved, at the expense of large pump values, while the often targeted low bias operation is strongly noise- and bandwidth-limited.  This fundamental investigation suggests that technological efforts should be oriented towards enabling large pump rates in nanolasers, whose performance promises to surpass microdevices in the same range of photon flux and input energy.
\end{abstract}


\maketitle 

\section{Introduction and objectives}

The development of semiconductor-based lasers has been the prime mover in device miniaturization, which, together with efficiency and cavity size reductions, has led to a steady lowering of the threshold pump value and an increase in efficiency, particularly since the conception of the Vertical Cavity Surface Emitting Laser (VCSEL)~\cite{Soda1979}.  The development of nanofabrication techniques have enabled the current realization of small cavities with high confinement and efficiency~\cite{Hill2014}, thus permitting the construction of coherent light sources with ultralow lasing thresholds.  Envisaged applications range from telecommunications, to spectroscopy, sensing, and probing of biological systems~\cite{Ma2019}.  

Focussing on applications related to the information society, one of the most attractive uses of nanolasers is the role they can play as light sources in optical interconnects~\cite{Miller2009,Tatum2015}.  These lie at the core of the operations performed in datacenters, the hubs which collect, store and handle the enormous amount of data generated and retrieved in today's World.  While playing a central role in the information society, datacenters have emerged as a major environmental challenge because of the amount of consumed power and of rejected heat~\cite{Lee2012,Whitehead2014,Whitehead2015}.  

The use of light, with its propagation speed at least two orders of magnitude larger than that of electrons, has promised great gains if electronics can be replaced by photonics, thus sparking a great deal of work in the development of optical chips~\cite{Smit2012,Notomi2014,Werner2017,Cheng2018,Soref2018}.  However, this grail cannot be reached without ever more efficient optical circuits and with low-consumption light sources~\cite{Tucker2011,Miller2017} to rein in energy waste in its different forms.

Within this context, nanolasers are expected to become the future sources of light in optical chips, thanks to their small footprint -- thus integration capabilities -- and low threshold, which promises considerable reductions in thermal load~\cite{Service2010}.  The direct encoding of information through laser modulation -- though replaceable by high-speed and low-power consumption modulators~\cite{Nozaki2017} -- is of course a highly desirable goal, as it simplifies architectures. As a consequence, the identification of the most suitable classes of devices, with their optimal operation conditions, and the assessment of their transmission potential is a current priority.

The question of speed enhancement in nanolasers has been debated for a long time.  Early warnings cautioned that even taking into account quantum electrodynamical effects in the estimate of the Purcell effect, nanolasers would not be able to outperform microdevices~\cite{Karlsson1994}.  Taking the opposite stance, extreme gain in laser response was predicted with transmission speeds largely in excess of 100 GHz~\cite{Altug2006,Sattar2015}.  Consideration of saturation effects~\cite{Suhr2010}, carrier dynamics~\cite{Lorke2010} and optimal design~\cite{Lau2009}, has concluded that the Purcell effect contributes in a more moderate way to extending the bandwidth of nanodevices and that some gain is to be expected at the nanoscale.  Even for metallic clad lasers initial enthousiastic claims~\cite{Shore2010,Ni2012} have been revised~\cite{Ding2015,Romeira2020} providing more realistic predictions.  

In essence, experimentally observed modulation speeds range from somewhat below 10 GHz up to 20 GHz, with devices built out of photonics crystals~\cite{Braive2009,Matsuo2010,Matsuo2011,Matsuo2014,Takiguchi2016} and nanowire lasers~\cite{Takiguchi2017}.  Few experimental results currently exist for metallic clad nanolasers, but the modulation values attained remain in the range cited above~\cite{Xu2019}, even though hopes appear to be high for much better bandwidths~\cite{Ding2015}.  Their main feature is a higher loss rate, introduced by the metallic cladding, which potentially offers better temporal response while requiring stronger pump because of the additional metal losses.  The required larger injection current values, however, introduce concerns about their practicality and, thus, about the actually achievable modulation bandwidths~\cite{Romeira2020}, due to potential damage issues~\cite{Khurgin2012,Khurgin2014}; these limitations may restrict the operation of these devices to pump ranges too close to threshold not to be strongly affected by noise~\cite{Wang2016,Pan2018,Moerk2018,Wang2020}.  Fano lasers offer an interesting alternative as they exploit a narrow resonance in a photonic--crystal--based platform~\cite{Moerk2014,Yu2014,Yu2015b}, with modulation speeds measured around 20 GHz~\cite{Yu2015,Yu2016}.  Micropillars, based on VCSEL technology, are the most advanced devices at this stage, with modulation bandwidths in the range up to 50 GHz~\cite{Westberg2012,Wolf2013,Haglund2015,Li2015,Feng2018,Shen2019}, reviewed in detail in~\cite{Feng2018}.  Only an injection locked-device (based on a photonic crystal) has surpassed them, at the cost of a more complex design, reaching a bandwidth $>$ 67 GHz~\cite{Chen2011}.  Ultralow consumption has been obtained with micropillars~\cite{Feng2018} and photonic crystals~\cite{Takeda2013}, while integration on chip has been recently achieved with a VCSEL structure~\cite{Liu2019}.

In addition to modeling specific features of devices, considerations on the potential of various classes of designs have been published (cf. e.g.,~\cite{Khurgin2014,Ding2015,Yu2015,Feng2018,Romeira2020}) and therefore will not be addressed here.  Our scope is a general investigation of the influence of noise on bandwidth, independently of the device platform and of construction details, thus outlining a methodology for the analysis of the potential impact of intrinsic fluctuations on the fidelity between laser output and modulation input.  We perform the investigation using a basic Quantum Well model~\cite{Coldren1995} since the procedure we establish is independent of the device's details.  In agreement with previous publications~\cite{Romeira2020}, our analysis quantitatively shows that high-quality reproduction of the input modulation can only be obtained sufficiently far from threshold, unless unusual schemes for encoding are taken into consideration~\cite{Wang2019B}.  Whether such bias values are practically attainable is a question left for technological discussions~\cite{Khurgin2012,Khurgin2014,Romeira2020}, but our results suggest that efforts oriented towards enabling large injection currents in nanolasers may well pay off the effort.

The paper is organized as follows:  model and numerical integration techniques are presented in Section~\ref{model}, followed by the numerical analysis, and its comparison with analytical predictions, of the free-running laser properties (Section~\ref{genres}).  A comparison of principle between the influence of noise on nano- and micro-lasers is offered in Section~\ref{adv}.  The numerical large-signal analysis is found in Section~\ref{large}, followed by conclusions (Section~\ref{concl}).

\section{Model and numerics}\label{model}

A standard rate equation model for nanolasers is used for the investigation~\cite{Bjork1991}, rewritten in a simplified form:
\begin{eqnarray}
\label{modeqS}
\dot{S} & = & -\Gamma_c S + \beta\gamma N(S+1) + F_S(t)\, , \\
\label{modeqN}
\dot{N} & = & P - \beta\gamma NS - \gamma N + F_N(t)\, ,
\end{eqnarray}   
where $S$ represents the photon number, $N$ the carrier number, $\Gamma_c$ the cavity inverse photon lifetime, $\gamma$ the carriers' relaxation rate, $\beta$ the fraction of spontaneous emission coupled into the lasing mode.  The dot over each variable ($\dot{S}$, $\dot{N}$) stands for the time derivative.  The product $\beta \gamma$ quantifies the amount of spontaneous emission coupled into the lasing mode and represents the coupling strength between photons and carriers.  $P$ is the pump rate which replenishes the carrier reservoir.  All carriers -- homogeneously broadened --  are taken in resonance with the cavity.  The noise terms $F_j$ are introduced using the Langevin hypothesis of gaussian-distributed (independent, random) events~\cite{Coldren1995} and satisfy the usual relation
\begin{eqnarray}
\langle F_i(t) F_j(t^{\prime}) \rangle & = & 2 F_{ij} \delta(t - t^{\prime}) \, , 
\end{eqnarray}
where $\delta$ is the Dirac delta distribution.

The explicit form of the noise coefficients as well as details of the implementation are given in~\cite{Lippi2018}.  The integration is performed using an Euler-Maruyama method, which allows for a simple, yet very effective way of obtaining good predictions of the model's noisy dynamics~\cite{Lippi2018}.  It is important to remark that this specific numerical implementation slightly overestimates the role of noise, without deviating too much from the expected analytical averages obtained directly from the rate equations.  Thus, for those parameter ranges for which information encoding appears to be feasible in the investigation, we can be quite confident in the result.  At the boundaries where the encoding capability may become marginal there may be a small quantitative shift.  However, since the investigation does not aim at obtaining specific predictions,  an eventual quantitative mismatch is of no relevance and does not impinge on the validity of the methodology presented in this study.

The integration of the rate equations with Langevin noise, eqs.~(\ref{modeqS},\ref{modeqN}), is not reliable close to threshold for a large $\beta$ laser, due to the imperfect reproduction of the threshold dynamics.  Comparison to numerical predictions obtained with fully stochastic processes (cf.~\cite{Roy2009,Roy2010,Vallet2019}, and more specifically~\cite{Puccioni2015} for a class B laser~\cite{Tredicce1985}) shows that the regime of strong spiking predicted by stochastic simulations -- and experimentally observed in mesoscopic devices~\cite{Wang2015,Wang2019} -- is entirely missed by the differential approach.  The origin of the discrepancy lies in the influence that the discreteness of the photon field has at low average values~\cite{Lebreton2013,Puccioni2015} and in the randomness of each event, which intrinsically introduces noisiness in the dynamics. 

The intrinsically very low photon number at threshold $\langle S \rangle_{th} \approx \beta^{-\frac{1}{2}}$~\cite{Rice1994}, which for large $\beta$ devices amounts to a few units, is a second element which prevents a correct differential description.  When considering a $\beta = 0.1$ nanolaser, $\langle S \rangle_{th}  \approx 3$, implying extremely large fluctuations, since the addition or removal of one photon represents over 30\% change.  Even though the photon number grows after threshold, its values remain small over a broad pump range, thus rendering the continuous description offered by the differential approach intrinsically faulty.  Finally, the small photon number makes it obvious that close to threshold the device cannot be operated to encode information!  

Since the range in which the dynamics is poorly described by the differential approach substantially overlaps the pump range where noise is too strong for reliable data transmission, we restrict our attention to pump values $\tilde{P} \ge 10$ (Fig.~\ref{IVcurve-g2}); here the use of rate equations is justified.  The existence of a broad region difficult to exploit for traditional data transmission -- i.e., encoding information in bits identified by different laser output power -- is not surprising in a $\beta = 0.1$ laser.  Indeed in these devices the transition between incoherent and coherent emission occupies a broad pump interval around threshold~\cite{Yokoyama1989,Bjork1991} and is characterized by a very noisy dynamics~\cite{Wang2016}, thus impeding the traditional data transmission.  Alternative information encoding techniques could be used in this region~\cite{Wang2019B}, but this topic is beyond the scope of this work. 

The parameter values which will be used throughout the investigation (unless otherwise specified) are:  $\Gamma_c = 10^{11} s^{-1}$, $\gamma = 3 \times 10^9 s^{-1}$, $\beta = 0.1$.  The pump will be normalized to a threshold value $P_{th}$, conventionally defined as its value for which the average photon number $\langle S \rangle_{th} = \beta^{-\frac{1}{2}}$~\cite{Rice1994}.  The use of the relative unit eases comparison among lasers with different parameter values, thus we introduce a normalized pump  defined as $\tilde{P} = \frac{P}{P_{th}}$.

All simulations are performed with a time step $t_s = 10^{-14} s$, which ensures a good reproduction of the dynamics~\cite{Lippi2018}.  In order to compact the information on the photon number, we bin the output on a timescale $t_d$ (simple addition of the photon over the timesteps in the $t_d$ window) in a way similar to what a photon detector would do.  The typical value used throughout the simulations is $t_d = 10^{-12} s$, but this number is checked for compatibility with the frequency $f_m$ at which the pump is modulated (cf. section~\ref{large}).  For a proper representation of the modulated dynamics $t_d \ll f_m^{-1}$; however, choosing a value of $t_d$ which is too small leads to an excess of temporal detail in the noisy trajectory, below physically meaningful timescales.  Thus, the quality of the simulations has been checked by changing $t_d$ to obtain a meaningful representation. 
All simulations have been done with programmes specifically written in C and in MatLab (including the routines).  Only the random number generators are standard:  for the MatLab programmes, we have used the internal libraries, for the C programmes those already specified in~\cite{Lippi2018}.  The predictions obtained from the two codes have been matched to ensure reproducibility.

\section{Unmodulated operation with noise -- comparison to analytical predictions}\label{genres}

Following traditional approaches~\cite{Lau2009,Shore2010}, we test the potential transmission bandwidth by simulating a sinusoidal modulation of the pump to infer the potential response of a $\beta = 0.1$ nanolaser.  Before considering the modulation, however, we study the laser behaviour with an unmodulated pump to evaluate the intrinsic noise influence.

\subsection{Dynamics and coherence of free running nanolasers}\label{free}

\begin{figure}[!t]
\centering
\includegraphics[width=1\linewidth]{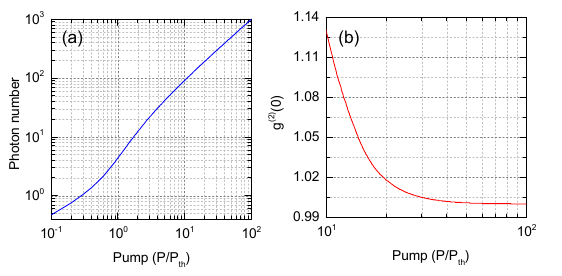}
\caption{Laser characteristic response (a) and second-order autocorrelation function $g^{(2)}(0)$ (b) of the free-running $\beta = 0.1$ laser.}
  \label{IVcurve-g2}
\end{figure}

\begin{table}
\begin{center}
\caption{Average photon number $\langle S \rangle$ and fluctuations $\sigma_S$ (standard deviation) for different pump values, normalized to pump threshold.  The last column gives the relative fluctuation amplitude.}
\setlength{\tabcolsep}{6mm}{
\begin{tabular}{|| c | c | c | c ||}\hline\hline
$\tilde{P}$ & $\langle S \rangle$ & $\sigma_S$ & $\frac{\sigma_S}{\langle S \rangle}$ \\ \hline\hline
10 & 91 & 37 &  0.41 \\ \hline
20 & 191 & 34 & 0.18 \\ \hline
50 & 491 & 30 & 0.06 \\ \hline
100 & 991 & 29 & 0.03 \\ \hline\hline
\end{tabular}}
\label{fluct}
\end{center}
\end{table}

\begin{figure*}[!t]
\centering
  \includegraphics[width=7.0in, height=3.2in]{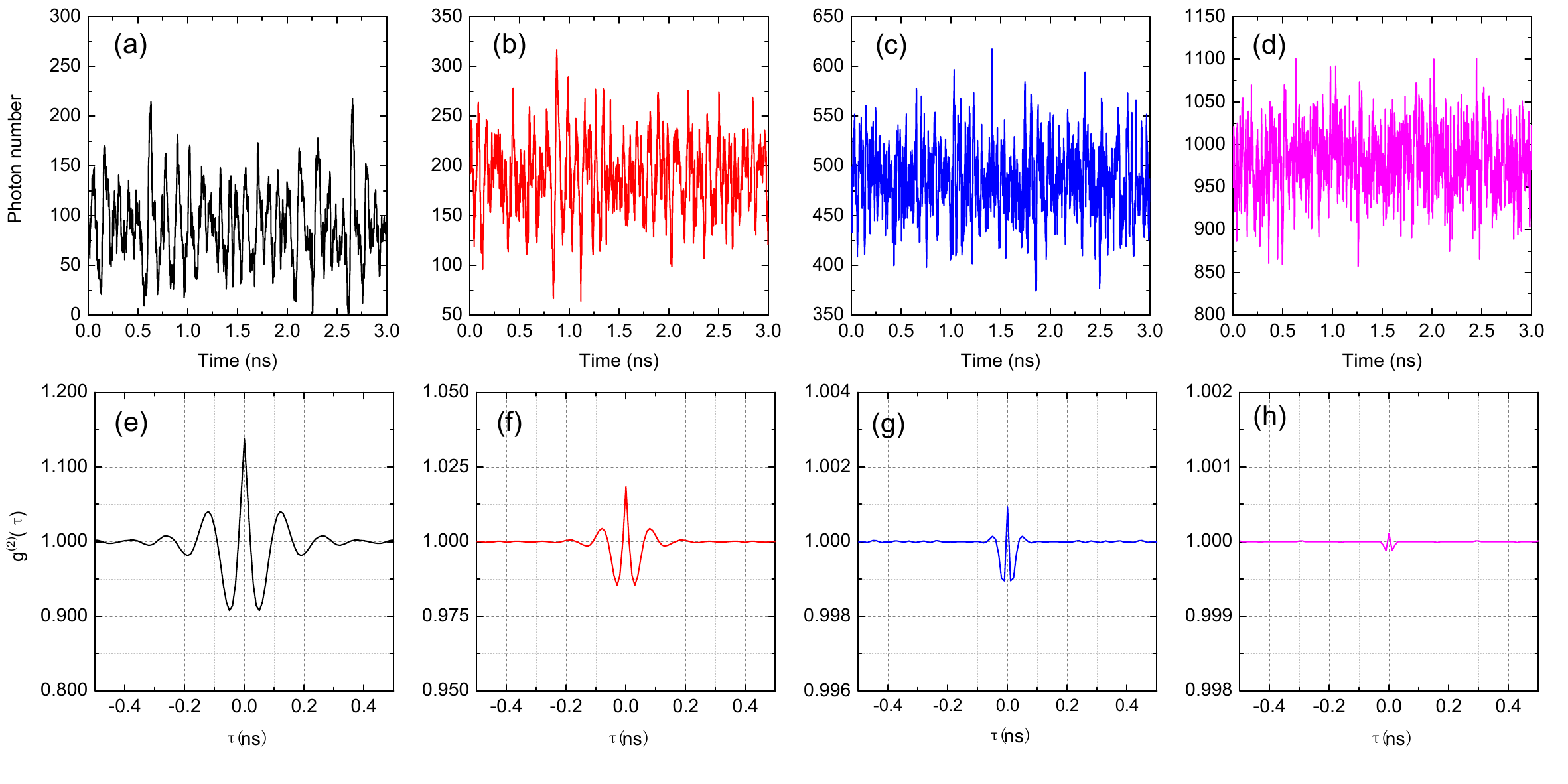}
  \caption{(a)-(d), Temporal dynamics of free running laser at $\tilde{P} =$ 10 (a), 20 (b), 50 (c) and 100, respectively; (e)-(h) Corresponding second-order time-delayed correlation function $g^{(2)}(\tau)$. }
  \label{correintime}
\end{figure*}

The integration of eqs.~(\ref{modeqS},\ref{modeqN}) provides dynamical information from which averages (e.g., input-output curves -- I/O --, Fig.\ref{IVcurve-g2}a) and the second-order autocorrelation function (Fig.\ref{IVcurve-g2}b) are obtained.  The weakly pronounced curvature of I/O response matches $\beta = 0.1$.  The second order, zero delay autocorrelation function (Fig.\ref{IVcurve-g2}b), computed from the temporal data sequence, is shown only in the interesting pump range and shows a rapid convergence towards Poisson statistics.  The fact that $g^{(2)}(0) > 1$ confirms that for $\tilde{P} \le 20$ noise is still sufficiently strong to substantially perturb the laser operation.

Quantitatively, noise influence is summarized in Table~\ref{fluct} through the ratio relative fluctuation size, substantial -- about 40\% -- at $\tilde{P} = 10$ and approaching 5\% only at $\tilde{P} = 50$.  In addition to confirming the shape of $g^{(2)}(0)$ (Fig.~\ref{IVcurve-g2}b), the table proves how noise must be fully taken into account when assessing data transmission.  

A direct illustration of the intrinsic noisiness of the laser output is given in Fig.~\ref{correintime}(a)-(d), which shows the photon number output by the laser in a $\Delta t = 3 ns$ time window at four pump values.  At $\tilde{P} = 10$ the (still quite) small photon number undergoes very large fluctuations (cf. also Table~\ref{fluct}) which progressively diminish, especially once $\tilde{P} = 50$ is reached (panel c).  The time-delayed, second order autocorrelation (Fig.~\ref{correintime}(e)-(h)) clearly shows that noise excites Relaxation Oscillations (ROs) (panels (e) and (f)), which leave only a hint of a half-period correlation at $\tilde{P} = 50$, to disappear entirely at the largest pump value considered.  Thus, the coupling between the relatively substantial noise amplitude and the small damping of the ROs gives rise to a robust intrinsic dynamics in the lower pump regions.  The influence of noise is directly illustrated by the strong variability in the oscillations amplitude at lower bias pump (Figs.~\ref{correintime}(a)-(b)).

\subsection{Comparison with analytical predictions}\label{an}

The results of section~\ref{free} can be compared to analytical predictions obtained from the linear stability analysis ({\it lsa})~\cite{Solari1996} of eqs.~(\ref{modeqS},\ref{modeqN}):

\begin{eqnarray}
\frac{d}{d t} 
\left(
\begin{array}{c}
\sigma \\
\nu \\
\end{array} \right) & = & \mathcal{\overline{\overline{T}}} \quad \left(
\begin{array}{c}
\sigma\\
\nu\\
\end{array} \right) \, , \\
\mathcal{\overline{\overline{T}}} & = & 
\left( 
\begin{array}{ c c }
\beta \gamma \overline{N} - \Gamma_c - \lambda & \beta \gamma (\overline{S} + 1) \\
-\beta \gamma \overline{N} & -\gamma - \beta \gamma \overline{S} - \lambda \\
\end{array} \right)
\end{eqnarray}
where we have defined the perturbations
\begin{eqnarray}
S(t) & = & \overline{S} + \sigma e^{\lambda t} \, , \\
N(t) & = & \overline{N} + \nu e^{\lambda t} \, ,
\end{eqnarray}
with the steady states for the photon and carrier number denoted by the overlines
\begin{eqnarray}
\label{Sss}
\overline{S} & = &  \beta^{-1} \left\{ \left(\frac{C-1}{2}\right) + \sqrt{ \left(\frac{C-1}{2}\right)^2 + \beta C} \right\} \, , \\
\label{Nss}
\overline{N} & = & \frac{\Gamma_c}{\beta \gamma} \frac{C}{1 + \beta \overline{S}}\, , \\
\label{normP}
C & = & \frac{P}{P_{th}}\, , \\ 
\label{Pth}
P_{th} & = & \frac{\Gamma_c}{\beta} \, .
\end{eqnarray}

The eigenvalues of $\overline{\overline{\mathcal{T}}}$ are:
\begin{eqnarray}
\nonumber
\lambda_{\pm} & = &  \frac{((\overline{S}-\overline{N})\beta + 1)\gamma+\Gamma_c}{2} \pm \\ \nonumber
& & \frac{1}{2} \left\{ \left[( \overline{S}^2 - 2 \overline{N} \, \overline{S} + \overline{N}^2 - 4 \overline{N}) \beta^2 + \right. \right.\\ \nonumber
& & \left. + (2 \overline{S} + 2 \overline{N})\beta+1\right] \gamma^2 + \\ 
\label{lambda+-}
& & \left. + ((-2 \Gamma_c \overline{S} - 2 \Gamma_c \overline{N}) \beta - 2 \Gamma_c) \gamma + \Gamma_c^2 \right\}^{\frac{1}{2}} \, .
\end{eqnarray}

\begin{figure}[htbp]
\centering
     \includegraphics[width=1\linewidth]{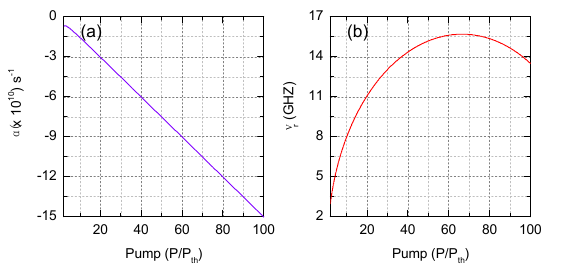}
\caption{
  Damping constant (a) and RO frequency $\nu_r =\frac{\omega_r}{2 \pi}$
(b) as a function of pump, computed from the {\it lsa}. 
}
  \label{damp-rosc}
\end{figure}

Above threshold, the eigenvalues, $\lambda_{\pm}$, of the linearized matrix $\mathcal{\overline{\overline{T}}}$ are (for most values of pump) complex conjugate and provide information on the RO frequency, as well as on the damping.  Writing them explicitly as
\begin{eqnarray}
\label{geneig}
\lambda_{\pm} & = & \alpha \pm i \omega_r \, , 
\end{eqnarray}
where $i$ is the imaginary unit, $\alpha$ the damping constant and $\omega_r$ the angular frequency of the RO, we can follow theie evolution in Fig.~\ref{damp-rosc}.  

Aside from a minimal reduction in its value extremely close to threshold -- too close to be of any relevance --, $\Re e \left\{ \lambda \right\}$ decreases steadily ($\alpha < 0, \, \forall \tilde{P} \ge 1$), thus showing a progressive stabilization of the nanolaser as the pump grows (Fig.~\ref{damp-rosc}(a)).  Comparing the values for $\tilde{P} = 10$ to those for $\tilde{P} = 100$, we recognize that the damping time constant is ten times larger for the higher pump.  This gives a first justification of the numerical observations of Fig.~\ref{correintime} (cf. also Table~\ref{fluct}) which show a reduction in the relative noise amplitude:  fluctuations are much more strongly damped as the pump is increased.

The RO frequency, $\nu_r = \frac{\omega_r}{2 \pi} = \frac{\Im m \left\{ \lambda \right\} }{2 \pi}$, shows a nonmonotonic behaviour, with a maximum close to $\tilde{P} = 65$, but a strong dependence on pump in the lower range (Fig.~\ref{damp-rosc}(b)).  From the figure, $\nu_r(\tilde{P}=10) = 8$ GHz, whose reciprocal provides a good match to the first maximum of the delayed autocorrelation (Fig.~\ref{correintime}(e)), representing the RO period.  The same holds for the comparison at the larger pump values, although the details become harder to discern.

\begin{figure}[htbp]
\centering
  \includegraphics[width=1\linewidth]{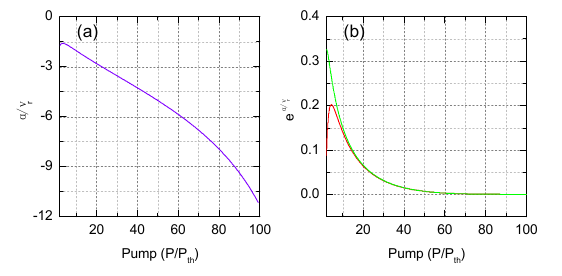}
    \caption{
  Relative damping, $\frac{\alpha}{\nu_r}$, (a) and attenuation of the oscillation computed over one period eq.~(\ref{att}) (b), as a function of pump, computed from the {\it lsa}.  The second curve (green) in panel (b) is computed for $\beta = 10^{-3}$.}
\label{atten}
\end{figure}

Additional information can be gained from a different representation. Considering the attenuation $\Delta S$ of the perturbed photon signal over time
\begin{eqnarray}
\Delta S \left( t = \frac{1}{\nu_r} \right)  & = & S \left( t = \frac{1}{\nu_r} \right) - \overline{S} \, , \\
\label{att}
& = & \sigma e^{\frac{\alpha}{\nu_r}} \, , 
\end{eqnarray}
we can plot the argument of the exponential (eq.~(\ref{att})), which represents the amount of damping in one oscillation period; in other words, we can see how much an excited RO is attenuated over its own period.  Fig.~\ref{atten}(a) depicts this relative damping $\left( \frac{\alpha}{\nu_r} \right)$, also showing a strong pump dependence.  Though weakest attenuation takes place around $\tilde{P} = 5$, at $\tilde{P} = 10$ it is nearly unchanged.  Until $\tilde{P} \approx 60$ the relative damping evolves linearly, to then double its reduction for $60 \le \tilde{P} \le 100$.

An even more telling indicator of (noise) damping is the residual perturbation amplitude at the RO period $T_r = \frac{1}{\nu_r}$, defined by the exponential function in eq.~(\ref{att}):  Fig.~\ref{atten}(b) shows a striking decrease, with pump, of the residual modulation at the end of a RO period.  While for $\tilde{P} = 10$ about 14\% of the oscillation amplitude remains, it decreases to 6\%, then 1\% at $\tilde{P} = 20$ and $50$, respectively (it is well below 1\% at $\tilde{P} = 100$).  
Thus, we see how for $\tilde{P} \lesssim 50$ the laser performs noisy ROs, while above $\tilde{P} = 50$ only a noisy trace remains (with noisy jumps, devoid of a specific frequency).  The strong damping  at large pump ensures that the continuous noise excitation is not capable of sustaining a RO.

In summary, the {\it lsa} shows that the laser's stability is poor in the range $1 \le \tilde{P} \lessapprox 20$.  Therefore it is not surprising that noise will wreak havoc in the laser's ability to encode information in that pump range.  These considerations match very well the autocorrelation results and give a strong foundation to the statement that one should not expect bias values close to threshold to be viable for useful information encoding and transmission.  These arguments add themselves to the consideration already made in section~\ref{model} concerning the small number of photons at threshold, which render the system extremely sensitive to photon noise (spontaneous emission), an element which cannot be taken into account by the {\it lsa}.

\begin{figure}[htbp]
\centering
  \includegraphics[width=1\linewidth]{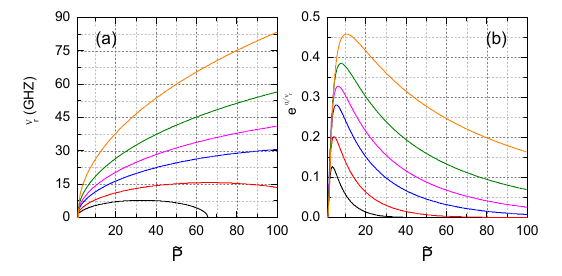}
    \caption{
    a)  Relaxation oscillation frequency as a function of (normalized) pump (as in Fig.~\ref{damp-rosc}b) for different values of cavity losses.  For comparison, $\Gamma_c = 1 \times 10^{11} s^{-1}$ is equivalent to a cavity quality factor $Q \approx 10^3$ (estimated on a micropillar device).  b)  Relative attenuation of the oscillation as in Fig.~\ref{atten}b.  $\Gamma_c = (0.5, 1, 2, 3, 5, 10) \times 10^{11}$ (black, red, blue, magenta, green and orange), respectively.
}
\label{gc}
\end{figure}

\subsection{Varying cavity losses}\label{cl}

While the material relaxation ($\gamma$) can vary only in a narrow range, the cavity losses ($\Gamma_c$) can span a range of at least one order of magnitude (and even more), at least when changing from one technology to another.  Typically, photonic crystal devices are manufactured for lower cavity losses (smaller $\Gamma_c$) while metallic-clad ones can withstand higher losses (micropillars are typically placed inbetween).  The change in $\Gamma_c$ is directly reflected into different modulation properties, in this model, and in the ensuing variations in relative damping.  In this section, we explore the consequences of varying $\Gamma_c$ to gain an insight into the advantages and disadvantages of each choice.  Again, only a detailed model tailored to specific devices will provide quantitative predictions.

Fig.~\ref{gc}a shows the evolution of the relaxation oscillation frequency as a function of pump ($\tilde{P}$) for different values of $\Gamma_c$.  As is well-known~\cite{Coldren1995}, $\nu_r$ increases with $\Gamma_c$, enabling a larger bandwidth and higher bias pump values.  However, accompanying the larger bandwidth a less efficient relaxation mechanism sets in (Fig.~\ref{gc}b), which strongly undamps any perturbation, leaving the device more vulnerable to noise.  This is particularly important when going towards metal-clad devices which, thanks to their larger intrinsic losses, are more amenable to faster modulation~\cite{Ding2015}.  Their technological limitations~\cite{Romeira2020}, which could restrict their operation to a few times above threshold, may position them in the most unfavourable regime as far as noise sensitivity is concerned.  Indeed,  the growth of the residual oscillation is very steep in the interval $1 \le \tilde{P} \le 10$, which corresponds exactly to the range which appears, for the moment, to be most easily accessible by realistic devices.  It is also important to recall that in the low-pump range noise may be much more damaging~\cite{Puccioni2015,Wang2016,Moerk2018,Pan2018,Romeira2020} than what can be forecasted by our differential analysis.  This prediction, though in need of verification through specific models and experimental measurements, raises serious concern about the possible usefulness of such devices for communications and counters the current enthusiasm surrounding metallic-clad nanolasers.

One observation which results from these general considerations, however, is that careful choice of the cavity losses can bring substantial benefits to the practical use of a device.  Comparing lasers with $\Gamma_c = 1 \times 10^{11} s^{-1}$ and $\Gamma_c = 2 \times 10^{11} s^{-1}$, we see that while the stability of the second at $\tilde{P} = 100$ is comparable to that of the first at $\tilde{P} = 50$ (the importance of this point will be clear in Section~\ref{large}), the modulation frequency attainable is (slightly) more than double as large at the maximum pump considered.  Thus, a shrewd choice of $\Gamma_c$ may bring considerable gain at a relatively low price, as long as technological limitations on pump density are lifted or, at least, pushed back.

Finally, it is important to remember that the curves are plotted as a function of a rescaled pump (normalized to threshold) for ease of comparison.  In physical pump units, they would appear shifted horizontally with respect to one another, due to the different amount of power needed to reach the individual threshold.

\begin{figure*}[!t]
\centering
  \includegraphics[width=7.0in, height=5.0in]{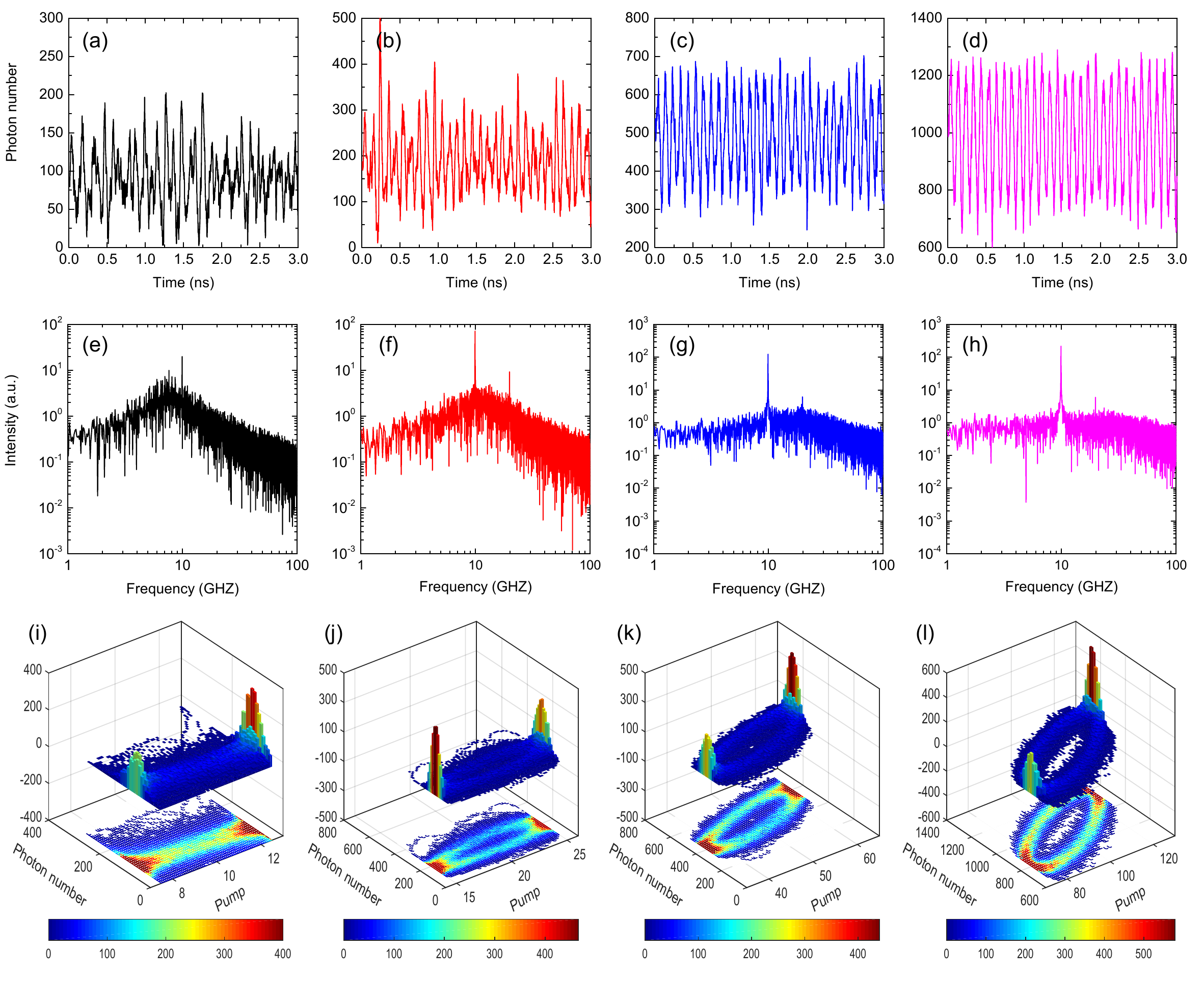}
  \caption{(a)-(d) Temporal dynamics of nanolaser modulated by 10 GHz large-signal at dc pump of 10P$_{th}$, 20P$_{th} $, 50P$_{th}$ and 100P$_{th}$, respectively; (e)-(h) the corresponding RF spectra (obtained by Fourier transform of the temporal signal -- also for the following figures); (i)-(l) the corresponding phase space responses.
  }
  \label{dynamics-10GHz-modulation}
\end{figure*}

\section{Nano- vs. microlasers}\label{adv}

The previous section suggests that data encoding ought to be most successful at very large pump values.  Leaving aside technological considerations~\cite{Khurgin2012}, the question arises naturally as to the existence of a true advantage in using a nanolaser instead of a microlaser.  Indeed, since threshold scales inversely with $\beta$ (eq.~(\ref{Pth})) and the output flux above threshold increases linearly with pump, introducing the shorthand notation $\beta = 10^{-n} \equiv \beta_{-n}$, we can easily establish the following relationship:  $\tilde{P}(\beta_{-n}) =  \frac{\beta_{-n}}{\beta_{-m}} \tilde{P}(\beta_{-m})$.  Setting $n=1$ and $m=3$, the threshold pump value for a microlaser with $\beta = 10^{-3}$ ($\tilde{P}(\beta_{-3})=1$) corresponds to $\tilde{P}(\beta_{-1}) = 100$, i.e., the maximum pump we have considered for our nanolaser.  Notice that the energy supplied to the microlaser is the same as the one needed for the nanolaser for the specified pumps (but not the pump density!).

Thus, the question is whether it would be more effective to modulate a microlaser twice above threshold to obtain reliable encoding, or a nanolaser pumped 200 times above threshold (the threshold value has to be avoided because of its intrinsic low stability).  Forgetting additional losses, the amount of power delivered to the devices is the same and the output flux the same.  Since from a technological point of view the microlaser is easier to operate, we need to see whether there are other features which may favour the nanodevice.

It is easy to see, by plotting $\Im m \left\{ \lambda_{\pm} \right\}$ (eq.~(\ref{lambda+-})), that $\nu_r$ is nearly the same for the two devices (not shown).  However, their damping coefficients differ.  This clearly appears from the relative damping (cf. Fig.~\ref{atten}b), weaker for $\beta_{-3}$ (green line).  Remembering that the graph is plotted in relative pump units, we need to compare the effective relative damping for the two lasers at pump values which differ by two orders of magnitude ($\tilde{P}(\beta_{-3}) = 100 \tilde{P}(\beta_{-1})$).  The far better stability of the nanolaser conclusively proves its superiority over the microdevice. This conclusion agrees with the Relative Intensity Noise results for the two devices at the respective pump values (cf. Fig. 3 in~\cite{Moerk2018}). We therefore conclude that our general analysis proves without doubt the preeminence of the smaller device, thus advocating for technological efforts in ensuring durable and reliable nanolaser operation at large pump.

\section{Dynamical response to large amplitude sinusoidal modulation}\label{large}

We now study the laser response to a large amplitude sinusoidal modulation to obtain information on the potential for information encoding~\cite{Morthier2000,Ding2015}.  The applied pump is modulated according to:
\begin{eqnarray}
\tilde{P}(t) = \tilde{P}_{dc} + \tilde{P}_m sin(2\pi f_m t)
\end{eqnarray}  
where $P_{dc}$ represents the dc bias, $P_m$ the modulation amplitude defined as a fixed fraction of the bias:  $P_m = m P_{dc}$, $m$ modulation depth of the sinusoidal signal.  $f_m$ is the modulation frequency and time is $t$.  On the basis of the considerations of section~\ref{genres} we choose $m = 0.25$ as the modulation depth to avoid coming too close to the pump interval where the laser is less stable, thus amplifying the influence of noise.  This value of $m$ is maintained throughout the simulations and amounts to a total modulation amplitude equal half the bias value.
The laser's response to this large amplitude modulation is going to be studied through two different parameter scans:  first we will fix the modulation frequency and look at the response as a function of bias pump, second, fixing the pump, we scan $f_m$.

\subsection{Fixed modulation frequency}\label{fixedfreq}

Fixing $f_m = 10$ GHz, we obtain the results of Fig.\ref{dynamics-10GHz-modulation}(a)-(d) for $\tilde{P}_{dc}$ = 10, 20, 50 and 100, respectively.  The response at $\tilde{P} = 10$ is very poor (Fig.\ref{dynamics-10GHz-modulation}(a)):  not only is the modulated photon flux very noisy, but careful inspection shows that the oscillations do not match a coherent sinusoid (frequent loss of phase/missing oscillations).  The picture evolves already at $\tilde{P} = 20$ since the modulation frequency is better reproduced, even though not perfectly; the amplitude, instead, remains extremely irregular.  For both bias choices, the laser stability is poor and the amount of intrinsic noise is too large to enable sufficient fidelity in signal reproduction (Table~\ref{minpump} gives the (normalized) pump extrema reached during modulation).

At $\tilde{P} = 50$ (Fig.\ref{dynamics-10GHz-modulation}(c)) the laser follows the pump modulation with a good quality signal, influenced, of course, by the residual amount of noise.  It is important to keep in mind that during the modulation the lower pump values reach $\tilde{P} = 37.5$, thus regions where both the stability and damping are less strong (Figs.~\ref{damp-rosc},~\ref{atten}); this results in a poorer stability and larger sensitivity to noise, as visible in Fig.~\ref{dynamics-10GHz-modulation}(c).  It is also the reason why at lower bias values the laser is unable to follow the modulation.  The largest pump values considered, $\tilde{P}_{dc} = 100$ (Fig.\ref{dynamics-10GHz-modulation}(d)) provides the best result both in terms of fidelity in modulation and in amplitude stability.

The power spectra offer complementary information.  At $\tilde{P}_{dc} = 10$ (Fig.~\ref{dynamics-10GHz-modulation}(e)) the broad RO is visible below the sharp peak of the modulation and matches the {\it lsa} predictions (Fig.~\ref{damp-rosc}). Comparison with this same figure shows that at $\tilde{P} = 20$ there is near coincidence between $f_m$ and $\nu_r$, while the ROs are nearly unrecognizable in the spectra at $\tilde{P}_{dc} = 50, 100$.  This is due to the fact that the damping is so strong for those pump values as to cancel all trace of the resonance.  A small harmonic component, indicating the presence of waveform distortion in the laser response, appears in the spectra of Figs.~\ref{dynamics-10GHz-modulation}(f)-(h), and is strongest for $\tilde{P}_{dc} = 20$, i.e., where the laser starts to follow but cannot entirely reproduce the waveform (cf. Fig.~\ref{dynamics-10GHz-modulation}(b)).  Notice that the frequency component at $f_m$ undergoes a non-monotonic evolution across the four pictures:  at $\tilde{P}_{dc} = 10$ it is smallest, due to the poor following of the pump modulation, it is highest at $\tilde{P}_{dc} = 20$ , thanks to the near-resonance between $f_m$ and $\nu_r$ (in spite of imperfect following).  Then it decreases in strength for $\tilde{P}_{dc} = 50$ , in spite of a better reproduction of the modulation -- however, the {\it lsa} shows that the RO frequency is much farther away at this pump value.  Indeed, the frequency component at $f_m$ gains strength again at $\tilde{P}_{dc} = 100$ due to a somewhat better match between $f_m$ and $\nu_r$.  In the absence of the results from the {\it lsa} (Section~\ref{an}), the power spectra alone would have not been sufficient to gain a good understanding of the unusual evolution of the forcing.  It is important to remark that the picture does not change if we choose a lower modulation frequency:  for sufficiently high bias values, the laser follows the driving  -- with even less dephasing -- while if $\tilde{P}_{dc}$ is too low, then the results are similar to those of Fig.~\ref{dynamics-10GHz-modulation}a (or worse).

\begin{figure*}[!t]
\centering
  \includegraphics[width=7.0in, height=5.0in]{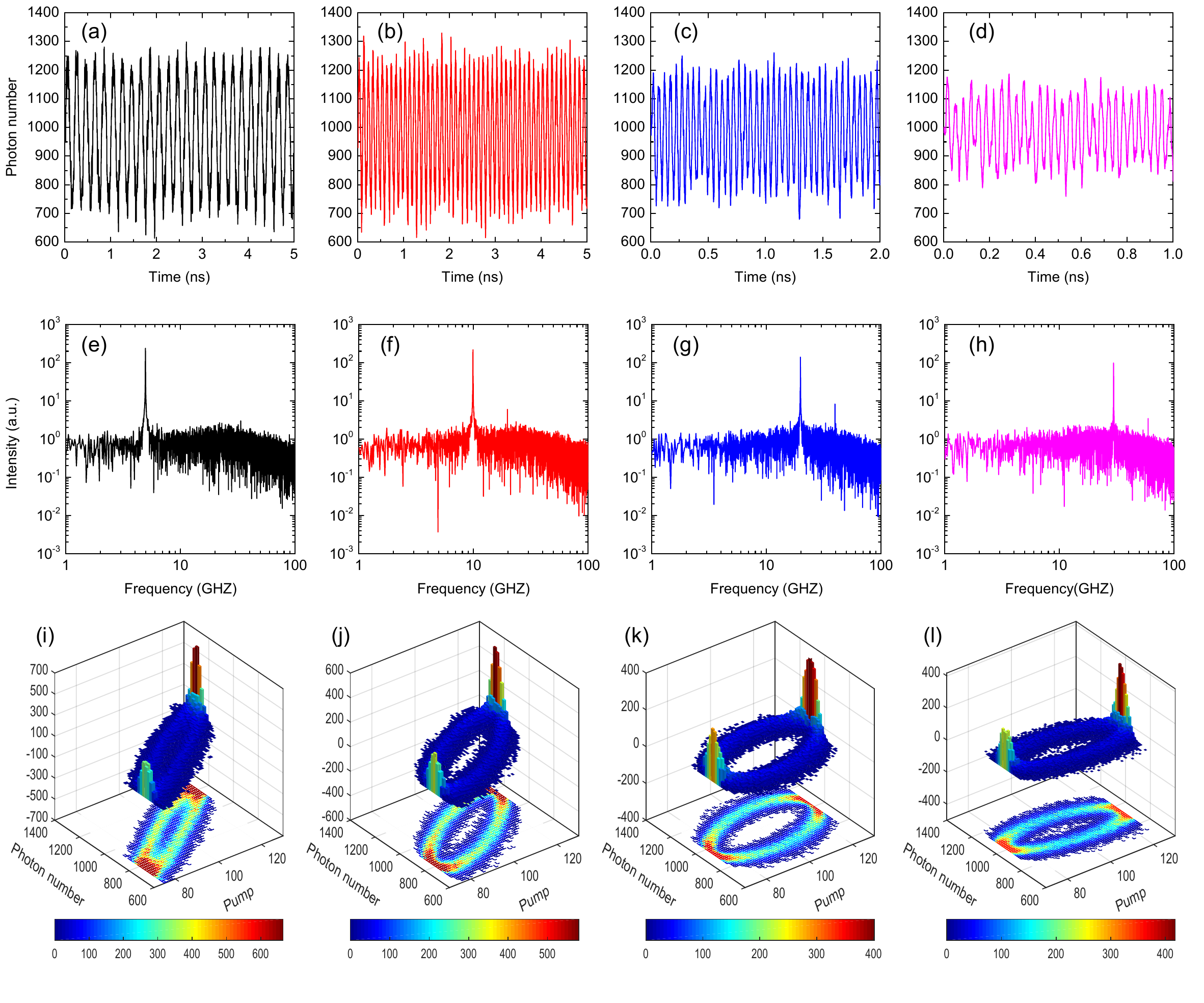}
  \caption{
  (a)-(d) Temporal dynamics of laser modulated at 100P$_{th}$ dc pump and by using 5 GHz, 10 GHz, 20 GHz and 30 GHz modulation frequency, respectively; (e)-(h) the corresponding RF spectra; (i)-(l) the corresponding phase space responses.}
  \label{100Pthdynamics}
\end{figure*}

\begin{table}
\begin{center}
\caption{Minimum $\tilde{P}_{min}$ and maximum $\tilde{P}_{max}$ pump values reached during modulation for the corresponding bias $\tilde{P}_{dc}$ (in normalized units).  Comparison to the results of the {\it lsa} provides useful information in understanding the modulation potential.}
\setlength{\tabcolsep}{8mm}{
\begin{tabular}{|| c | c | c ||}\hline\hline
$\tilde{P}_{dc}$ & $\tilde{P}_{min}$ & $\tilde{P}_{max}$ \\ \hline\hline
10 & 7.5 & 12.5 \\ \hline
20 & 15 & 25 \\ \hline
50 & 37.5 & 62.5 \\ \hline
100 & 75 & 125 \\ \hline\hline
\end{tabular}}
\label{minpump}
\end{center}
\end{table}

Another representation is offered by the phase space reconstruction (Figs.~\ref{dynamics-10GHz-modulation}(i)-(l)) where the frequency of the photon number occurrences is plotted in the plane of photon number and normalized pump.  The absence of a proper oscillation in the photon number, $S$, remarked for $\tilde{P}_{dc} = 10$ translates into a graph (Figs.~\ref{dynamics-10GHz-modulation}(i)) where, aside from the two peaks which correspond to maximum and minimum of the oscillation (and therefore occur more frequently), there is not much of a recognizable structure.  The 2D projection of the distribution (bottom plane) gives another illustration of the previous statement:  there is no visible structure in the frequency distribution of the observed emission levels, thus implying a lack of a usable relationship between input and output.

As the bias pump grows, a structure emerges, where the laser's ability in following the pump modulation is represented by the ellipse of points (cf. also 2D projection).  At  $\tilde{P}_{dc} = 20$ a certain amount of scatter is still observable, while it disappears almost entirely for the two larger pump values.  The opening of the ellipse signals the onset of a dephasing between pump modulation (which oscillates along the right axis) and laser response (left axis).  The increasing phase lag is probably due to the larger modulation amplitude, as $\tilde{P}_{dc}$ grows.

\subsection{Fixed modulation amplitude}\label{fixedampl}

Complementary information is provided by a scan in modulation frequency.  Given the optimal response offered by the largest bias value ($\tilde{P}_{dc} = 100$), we choose it as the reference bias, together with the fixed modulation amplitude which follows.  Figs.~\ref{100Pthdynamics}(a)-(d) shows the laser response as a function of modulation frequency.  At $f_m = 5 GHz$ and $10$ GHz the modulation is very well reproduced, with an amplitude in the response which matches the modulation depth $\left( \frac{S_{max} - S_{min}}{S_{dc}} \approx \frac{\tilde{P}_{max} - \tilde{P}_{min}}{\tilde{P}_{dc}} \right)$.  At $f_m = 20$ GHz a partial attenuation of the response is perceptible, while it becomes strongly visible at $f_m = 30$ GHz, where the photon modulation has lost more than half of its amplitude.  Nonetheless, the modulation signal is reproduced with good fidelity, albeit with an increasingly noisy amplitude.

The power spectra (Figs.~\ref{100Pthdynamics}(e)-(h)) do not add much information to this picture:  a good quality signal, more than 40 dB above the background, appears in Fig.~\ref{100Pthdynamics}(e), with the beginning of a harmonic component in Figs.~\ref{100Pthdynamics}(f).  As $f_m > \nu_r$ (Fig.~\ref{100Pthdynamics}(g)), the amplitude decreases while the second harmonic grows somewhat.  The strength of the response at $f_m$ remains, however, close to the 40 dB level.  Instead, the peak narrows considerably at $f_m = 30$ GHz and loses some more contrast.  However, the peak is well beyond 20 dB even in this case, signalling the potential for modulation even at more than double the RO frequency.

\begin{figure}[!t]
\centering
  \includegraphics[width=1\columnwidth]{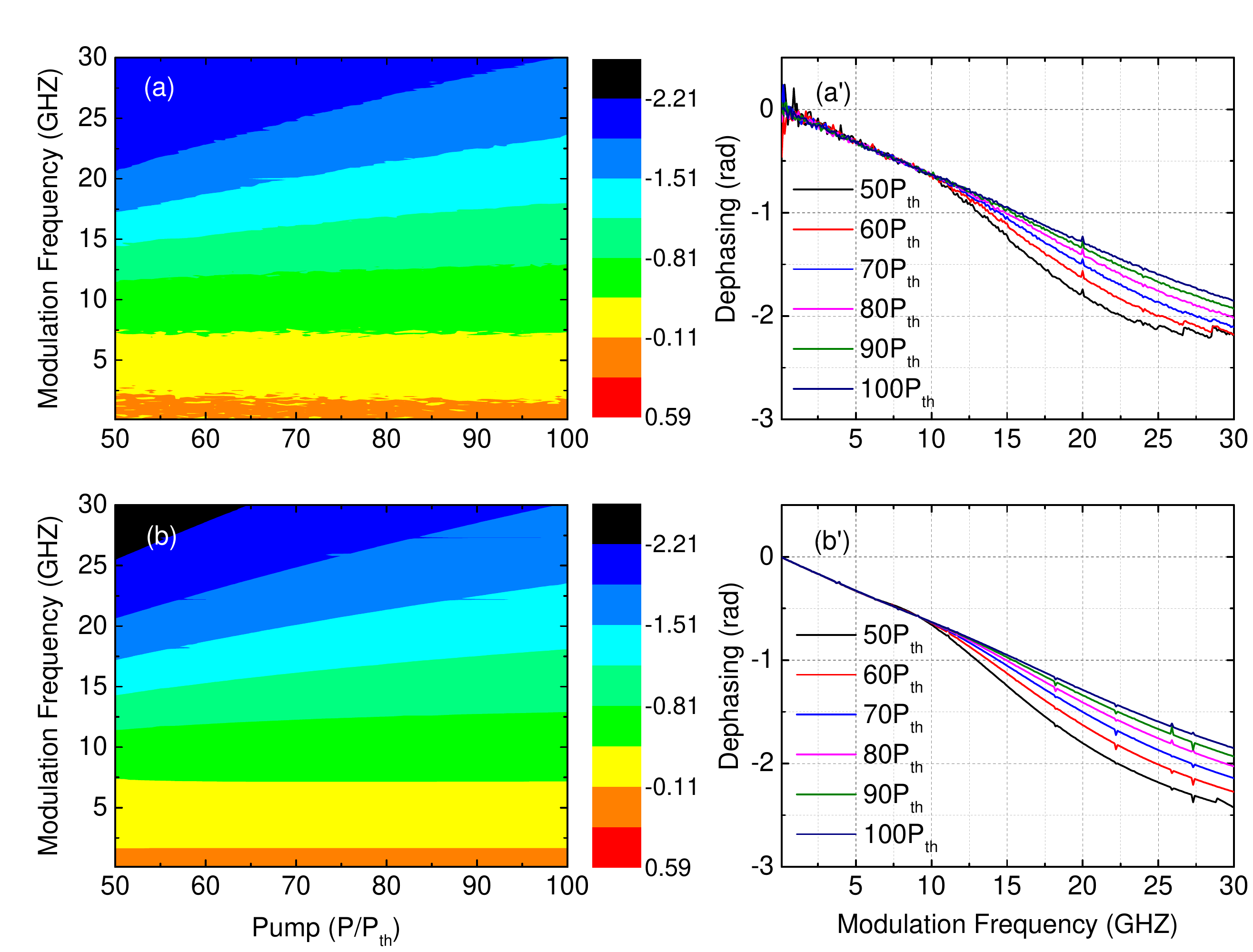}
  \caption{
  Dephasing contour plot in the plane of modulation frequency and dc pump: (a) and (a’) with noise; (b) and (b’) without noise; (a’) and (b’) are the dephasing function curves as the function of modulation frequency.}
  \label{dephasing}
\end{figure}

The phase space plots (Figs.~\ref{100Pthdynamics}(i)-(l)) convey similar information to what we have seen in Fig.~\ref{dynamics-10GHz-modulation}(i)-(l):  at low frequency the laser output follows nearly instantaneously the modulation, while a delay develops as the frequency grows and manifests itself both in a rotation of the major axis of the ellipse and in the opening of the ellipse itself.  This representation, however, confirms the viability of principle of a modulation up to 30 GHz for $\tilde{P}_{dc} = 100$.  These results seems to indicate that, at least for a sinusoidal modulation, the -3dB bandwidth empirical rule, which estimates the maximum modulation frequency at $\sim 1.3 \times \nu_r$~\cite{Tucker1986}, is quite conservative, since the current predictions give a prefactor $2$ (instead of $1.3$).

As the modulation frequency increases the phase lag, which accompanies the crossing of the RO resonance and is ultimately responsible for the laser's inability to follow the modulation, develops between the driving and the laser response.  Fig.~\ref{dephasing} quantifies the influence of noise on dephasing, computed for the set of parameters of Fig.~\ref{100Pthdynamics}.  Panels (b) and (b') characterize the average dephasing in the absence of noise, where
\begin{eqnarray}
\langle \delta \phi \rangle = \left\langle \arg \left\{ \frac{\vec{S}(t)}{|S(t)|} - \frac{\vec{P}(t)}{|P(t)|} \right\} \right\rangle_t \, ,
\end{eqnarray}
where the brackets $\langle \cdot \rangle_t$ represent the ensemble average obtained from the sequence of temporal predictions.  Panel (b), colour-encoding $\langle \delta \phi \rangle$, shows that for modulation frequencies up to $\sim$ 8GHz the dephasing is nearly independent of the bias pump value; for larger modulation frequencies, the influence of $\tilde{P}$ increases, showing a more marked increase in $\langle \delta \phi \rangle$ in the lower bias range.  This explains the increased difficulty the nanolaser experiences in following the pump modulation as the frequency grows.  Panel (b') gives a more quantitative picture of $\langle \delta \phi \rangle$ as a function of $f_m$ for the set of pump values indicated.  There appears a marked difference (about 20\%) with larger dephasing for $\tilde{P} = 50$ than for $\tilde{P} = 100$, with $\langle \delta \phi \rangle \approx \pi$ reached at $f_m \approx 25 GHz$ for $\tilde{P} = 100$ (and at $f_m \approx 18 GHz$ for $\tilde{P} = 50$).

Noise (panels (a) and (a')) lessens somewhat the deterministic average dephasing in the lower pump range, reducing the spread in $\langle \delta \phi \rangle$ at $f_m = 30 GHz$.  Other than in the upper frequency range, for the lower bias values considered, the influence of noise does not qualitatively change the deterministic behaviour of the average dephasing, which governs the ability of the nanolaser to follow the pump dynamics.

Complementary information on the nanolaser's ability to follow the input signal can be gathered by computing the following indicator:
\begin{eqnarray}
\delta S_{rms} = \sqrt{ \frac{1}{N} \sum_{i=1}^N \frac{ \left( S_{i,n} - S_{i,d} \right)^2}{S_{i,d}^2}} \, ,
\end{eqnarray}
i.e., the relative root-mean-square ({\it rms}) deviation in the photon number, where $N$ is the number of computed points in a temporal evolution of the driven laser output, $S_{i,n}$ is the $i$-th computed photon number in the presence of noise and $S_{i,d}$ the corresponding value computed from the deterministic trajectory.  The computation of the trajectory without noise ensures the same phase delay, induced by the phase lag, and thus enables the quantitative assessment of the influence of noise on the photon output.  This indicator provides therefore a complementary piece of information which helps quantifying the laser performance.

\begin{figure}[!t]
\centering
  \includegraphics[width=1\columnwidth]{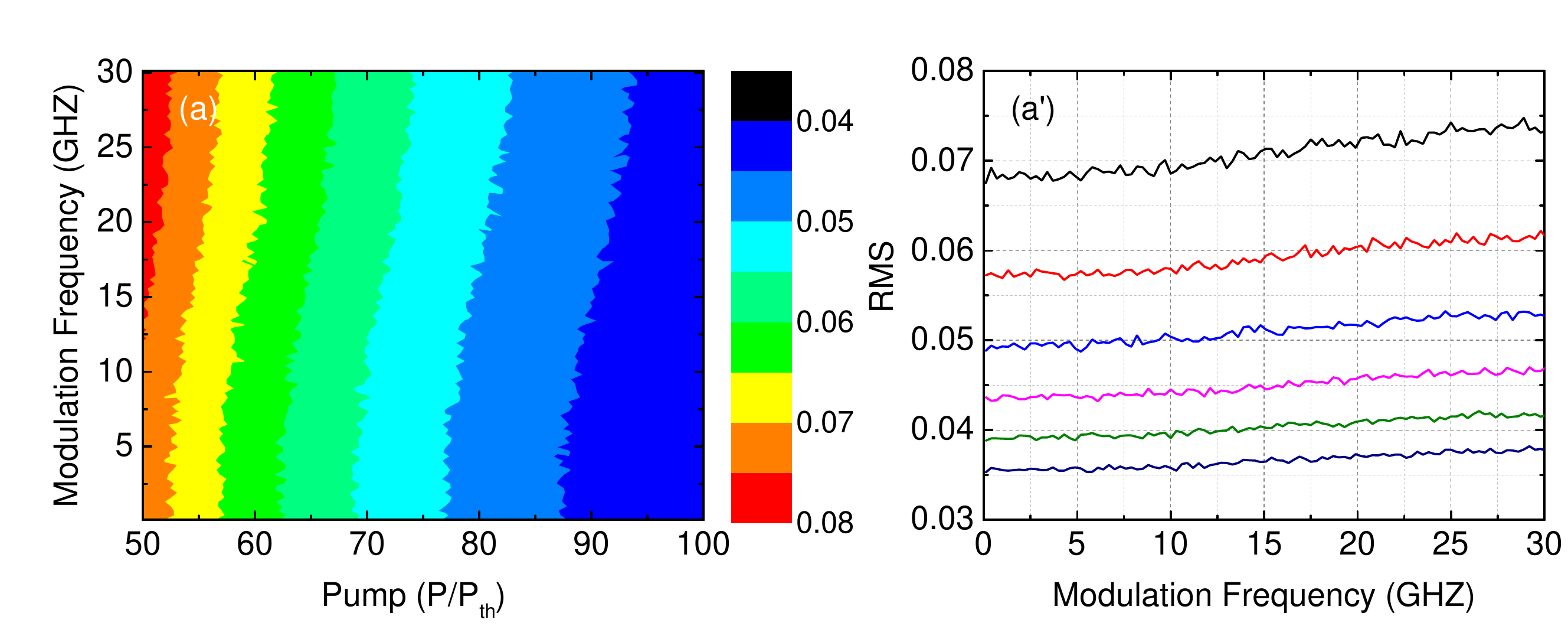}
  \caption{
  (a) relative noise contribution (colour-encoded) as a function of modulation frequency (vertical) and normalized pump (horizontal). (a') evolution of $S_{rms}$ at constant pump for $\tilde{P} =$ 50, 60, 70, 80, 90, 100 (from top to bottom), respectively.
  }
  \label{rms}
\end{figure}

Fig.~\ref{rms} shows a synopsys of the observations.  In the left panel, $\delta S_{rms}$ is colour encoded (cf. bar on the right of the figure) and is plotted as a function of normalized pump (horizontal axis) and modulation frequency (vertical axis). The overall trend is a reduction in the relative photon  fluctuation as the pump increases:  the value of $\delta S_{rms}$ decreases with increasing pump.  On the other hand, a moderate increase in $\delta S_{rms}$ is visible when the modulation frequency grows.  This point is stressed by the graph on the right panel which shows the evolution of the {\it rms} noise contribution as a function of modulation frequency for given pump values:  the influence of pump is clearly visible, since doubling the pump (from 50 to 100) reduces by a factor 2 the $\delta S_{rms}$ (cf. figure caption for details).  On the other hand, the frequency contribution is rather marginal, as all curves are sensibly horizontal.  It is important to notice that the graph is plotted starting from $\tilde{P} = 50$, since below the quality of the signal is too poor to warrant the use of $\delta S_{rms}$ (Fig.~\ref{dynamics-10GHz-modulation}a,b).  The same holds for the computation of the dephasing (Fig.~\ref{dephasing}).  The indicator confirms what could be expected from the {\it lsa} (in principle valid only for small perturbations):  the growing relative damping (Fig.~\ref{atten}), as a function of pump, improves the noise performance.

\section{Conclusions}\label{concl}

Our numerical investigation on the performance of a nanolaser subject to a large-signal modulation of its pump has been aimed at identifying the role that intrinsic noise plays in the response.  Its target has been the establishment of a methodological procedure, which we have applied to a model capturing the main features of a small-sized semiconductor-based device, independently of its technological features (photonic crystal, micropillar, metal-clad, etc.).  The investigation method is based on analytical results which hold for small-signal modulation and which provide insight into the main physical features responsible for the laser's noise sensitivity.  This procedure can be repeated for specific laser models to obtain specific and accurate estimates of their performance by taking into account the technological details of each single device.

In addition to the establishment of a methodology, the other notable result of the investigation consists in the identification of the parameter range (bias pump and modulation frequency) in which the pump modulation is satisfactorily followed by the nanolaser.  The results indicate that large bias is indispensable for good signal reproduction.  While technological restrictions exist, this analysis suggests that efforts to overcome them would be strongly beneficial for the use of nanolasers in the telecommunications.  Comparing the nanolaser results to the performance expected from a microlaser {\it under the same input power and photon flux} -- thus, in a regime for which no technological problems are expected from the larger device -- shows that the nanodevice largely outperforms its lower $\beta$ counterpart.  This is in agreement with the features of the Relative Intensity Noise and clearly appears from our general analysis.  Indicators introduced to quantify the average detuning and the relative  deviations in the photon number introduced by noise confirm the benefits of a large bias choice.

\begin{acknowledgments}
T.W. acknowledges the financial support from the National Natural Science Foundation of China (Grant No. 61804036), Zhejiang Province Commonweal Project (Grant No. LGJ20A040001), G.W. acknowledges National Key R \& D Program Grant (Grant No. 2018YFE0120000), and Zhejiang Provincial Key Research and Development Project Grant (Grant No. 2019C04003).
\end{acknowledgments}


\end{document}